\documentclass[runningheads]{llncs}

\usepackage{graphicx}
\usepackage{subcaption}
\usepackage{derivative}
\usepackage[hidelinks]{hyperref}
\usepackage{xurl}

\begin{document}
\title{Can Interpretability Layouts Influence Human Perception of Offensive Sentences?\thanks{This research was supported by the EU funded VALAWAI (\#~101070930) and WeNet (\#~823783) projects, the Spanish funded VAE (\#~TED2021-131295B\-C31) and Rhymas (\#~PID2020-113594RB-100) projects, and the Generalitat de Catalunya funded \emph{Ajuts a grups de recerca de Catalunya} (\#~2021 SGR 00754) project.}}

\titlerunning{Interpretability Layouts and Human Perception}
%
\author{Thiago Freitas dos Santos\inst{1,2}\orcidID{0000-0003-3800-7243} \and
Nardine Osman\inst{1}\orcidID{0000-0002-2766-3475} \and
Marco Schorlemmer\inst{1}\orcidID{0000-0002-9591-3325}}
\authorrunning{T. Freitas dos Santos et al.}
%
\institute{Artificial Intelligence Research Institute (IIIA), CSIC, Barcelona, Catalonia, Spain \and
Universitat Autònoma de Barcelona, Catalonia, Spain
\email{\{thiago,nardine,marco\}@iiia.csic.es}}
\maketitle              

\begin{abstract}
This paper conducts a user study to assess whether three machine learning (ML) interpretability layouts can influence participants' views when evaluating sentences containing hate speech, focusing on the ``Misogyny'' and ``Racism'' classes. Given the existence of divergent conclusions in the literature, we provide statistical and qualitative analyses of questionnaire responses using the Generalized Additive Model to estimate participants' ratings, incorporating within-subject and between-subject designs. While our statistical analysis indicates that none of the interpretability layouts significantly influences participants' views, our qualitative analysis demonstrates the advantages of ML interpretability: 1) triggering participants to provide corrective feedback in case of discrepancies between their views and the model, and 2) providing insights to evaluate a model's behavior beyond traditional performance metrics.

\keywords{Interpretability  \and User study \and Online interaction.}
\end{abstract}

\section{Introduction}
\label{sec:Intro}

Online communities establish norms to regulate interactions between agents (community members) with diverse backgrounds and views. A normative system that intends to regulate community members' behavior faces the challenge of continuously learning what constitutes a norm violation as communities' views evolve (e.g., change in what characterizes a violation, emergence of new violation classes). To address this challenge, our previous work~\cite{freitas2023cross} proposed a Machine Learning (ML) framework that supports normative systems to continuously learn what constitutes a norm violation from community members' feedback. This framework comprises transformer-based models~\cite{vaswani2017attention}. Specifically, it combines adapters with Pre-Trained Language Models~\cite{houlsby2019parameter}, an efficient neural network architecture for addressing Natural Language Processing (NLP) tasks~\cite{min2021recent}. We evaluated this framework by investigating the use case of article editing in Wikipedia~\cite{freitas2023cross}, an online platform where diverse people connect in an open environment~\cite{wiki}. In this context, actions are community members' attempts to edit articles, and we focused on the ``no hate speech'' norm (the requirement to not engage in hate speech), considering edits that are categorized into two hate speech classes: ``Racism'' and ``Misogyny.''\footnote{Disclaimer: This paper presents offensive language that may disturb some audiences.}

We argued in~\cite{freitas2023cross} that, in addition to identifying violations of the “no hate speech” norm, normative systems must explain the different views manifested in online communities, providing information on which elements of an action (words of a sentence) contribute to the model's output. To address this challenge, we employed ML interpretability using the Integrated Gradients (IG, Section~\ref{subSec:IG}) algorithm, which constitutes a key contribution of~\cite{freitas2023cross}. This approach aligns with the principles of responsible artificial intelligence, promoting transparency and facilitating community members' comprehension of what constitutes hate speech. It also supports model debugging and creates the conditions for triggering collaborative feedback elicitation when community members agree there are discrepancies between the model's output and their views on hate speech. 

\begin{figure*}[t!]
\captionsetup{font=small}
\centering
\begin{subfigure}{0.50\textwidth}
  \includegraphics[width=1\linewidth]{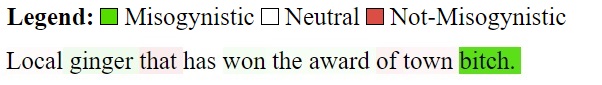}
  \caption{Local interpretability.}
  \label{fig:UserStudyLocalInterpretability}
\end{subfigure}
\begin{subfigure}{0.42\textwidth}
  \includegraphics[width=1\linewidth]{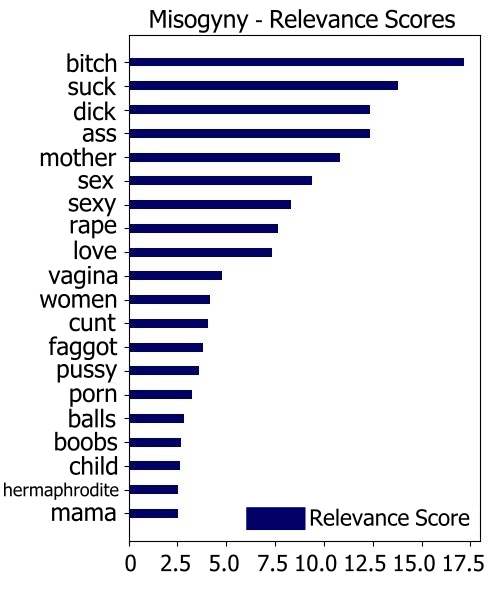}
  \caption{sum of relevance scores.}
  \label{fig:UserStudySumRelevance}
\end{subfigure}

\caption{The local interpretability and sum of relevance scores layouts. Figure~\ref{fig:UserStudyLocalInterpretability} represents the interpretability of classifying a sentence with the ``Misogyny'' class, while Figure~\ref{fig:UserStudySumRelevance} illustrates the sum of relevance scores for the ``Misogyny'' class considering words across all instances in the training dataset. In Figure~\ref{fig:UserStudyLocalInterpretability}, the intensity of the green shade indicates the relevance of the highlighted word to ``Misogyny,'' demonstrating the contribution of those words to identifying the text as misogynistic. In contrast, the intensity of the red shade is related to the decrease in the violation confidence, demonstrating the contribution of those words to classifying the text as not misogynistic.}
\label{fig:UserStudyAllSettings}
\end{figure*}

The IG algorithm enabled our framework~\cite{freitas2023cross} to present interpretability information to community members in three layouts (Figure~\ref{fig:UserStudyAllSettings}). First, Figure~\ref{fig:UserStudyLocalInterpretability} depicts an example of the local interpretability layout. It describes the impact of each word on identifying a given text as misogynistic, enabling users to understand specific problematic words in their text that may contribute to this classification. Second, Figure~\ref{fig:UserStudySumRelevance} presents the sum of the relevance scores, which is obtained by summing the relevance value of each word based on its occurrence in the entire dataset used to train the ML model (described in~\cite{freitas2023cross}). This layout provides users with a comprehensive general overview of problematic words that may result in misogynistic classification, representing the community's current view. The third layout combines Figures~\ref{fig:UserStudyLocalInterpretability} and~\ref{fig:UserStudySumRelevance}. It first presents the local interpretability, followed by the list of the most relevant words.  

Thus, as the layouts in Figure~\ref{fig:UserStudyAllSettings} present different interpretability information to community members, the main contribution of this work is to conduct a user study to investigate whether interpretability layouts can influence participants' views when evaluating sentences containing hate speech (Section~\ref{sec:Design}), providing empirical evidence on the effective use of ML interpretability~\cite{doshi2017towards}. This is especially relevant considering that the existing literature exhibits divergent conclusions on the impact of ML interpretability~\cite{hase2020evaluating,rong2022towards} (Section~\ref{sec:Literature}). For instance, some works suggest that how interpretability information is presented can influence people in different tasks (e.g., understanding recommender systems~\cite{radensky2022exploring}, lowering an ML model classification probability~\cite{arora2022explain}). In contrast, other works indicate that interpretability information does not significantly affect how people execute tasks (e.g., simulate a model's behavior on new reviews~\cite{arora2022explain}, predict age~\cite{chu2020visual}). 

In our experiment, participants answered an online questionnaire to choose whether they agreed or disagreed with classifying a sentence as ``Racism'' or ``Misogyny'' using a 7-point Likert scale. We employ the Generalized Additive Model (GAM, Section~\ref{subSec:GAM}) to estimate the participants' ratings, considering within-subject and between-subject designs. The statistical results indicate that none of the interpretability layouts significantly influence participants' views regarding the classification of hate speech (Sections~\ref{subSec:Within} and~\ref{subSec:Between}). Despite that, our qualitative analyses contribute valuable insights regarding participants' familiarity with hate speech evaluation, questionnaire design, and the presence of discrepancies between the ML model and participants' views (Section~\ref{subSec:Quali}). This allows us to understand the impact of incorporating interpretability in ML systems. First, interpretability can enhance people's comprehension of words relevant to the ML model's output, which triggers them to provide corrective feedback in cases of discrepancies. Second, it provides insights into evaluating the ML model's behavior beyond traditional performance metrics (e.g., recall, F1-score, and precision), which is important when understanding the rationale behind an ML model's decision is essential for optimal implementation. 

\section{Background}
\label{sec:Back}
This chapter presents the tools used in this work. First, we detail Integrated Gradients (IG), the interpretability algorithm. Then, we present the Generalized Additive Model (GAM) used to analyze the user responses. 

\subsection{Integrated Gradients (IG)}
\label{subSec:IG}
IG obtains the relevant words associated with hate speech classification, enabling our solution to gain insights into the inner workings of transformer-based models~\cite{freitas2023cross}. It accomplishes that by debugging and extracting rules from a Deep Neural Network (DNN)~\cite{sundararajan2017axiomatic}. Specifically, IG calculates a word's contribution to detecting hate speech by a backward pass through the model, propagating its relevance from the output to the input~\cite{lyu2022towards}. The central assumption of this algorithm is that the tokens with the highest gradient values present the most substantial influence on the classification output. 

Following the formalization in~\cite{sundararajan2017axiomatic,lyu2022towards} and considering an NLP task, let $x$ be the sentence formed by a set of tokens $x_i,i \in {1,2, ...n}$ and $\bar{x}$ the baseline input represented by a zero embedding vector. $\pdv{M(x)}{x_i}$ is the gradient for token $i$ and $M$ is the transformer-based model. Theoretically, to obtain the integrated gradients, IG considers a straight-line path from the baseline $\bar{x}$ to the input $x$, computing the gradients at all points of the path~\cite{sundararajan2017axiomatic}. Thus, the integrated gradients come from the accumulation of these individual points. Equation~\ref{eq:IntegratedGradsIntModel} formalizes the integral calculation. 

\begin{equation} \label{eq:IntegratedGradsIntModel}
	IntGrads(x_i) \gets (x_i - \bar{x_i}) \odot \int_{\alpha =0}^{1} \pdv{M \times (\bar{x} + \alpha \times (x - \bar{x}))}{x_i} \times d\alpha
\end{equation}

However, to efficiently compute the integrated gradients, IG approximates $IntGrad(x_i)$ by the Riemann sum method (Equation~\ref{eq:IntegratedGradsScore}), which defines a set of finite points ($m$) along the straight-line path. $r(x_i)$ is the calculated relevance score and $m$ is chosen empirically. Experiments in~\cite{sundararajan2017axiomatic} suggest around 20-300 points along the path.

\begin{equation} \label{eq:IntegratedGradsScore}
    r(x_i) \gets (x_i - \bar{x_i}) \odot \sum_{k=1}^{m} \pdv{M (\bar{x} + \frac{k}{m} \times (x - \bar{x}))}{x_i} \times \frac{1}{m}
\end{equation}

\subsection{Generalized Additive Model (GAM)}
\label{subSec:GAM}

The analysis of response data in our user study requires the consideration of four factors that may influence participants' views: the interpretability layout, the demographic information, the participants, and the sentences being evaluated (Section~\ref{sec:Design}). Following this requirement, we adopt the Generalized Additive Model (GAM) that enables us to explore the relationship between distinct variables by using nonlinear functions to model the response data~\cite{fahrmeir2013regression}. Thus, by capturing the relationship between these variables with a GAM, we can estimate participants' ratings about hate speech, discovering whether any of these factors influence participants' views.

Following the formalization of~\cite{hastie1987generalized} that is described in Equation~\ref{eq:GAM}, GAMs comprise a sum of smooth functions that can embody linear effects and variables, including continuous and categorical. This composition effectively allows GAMs to capture nonlinear relationships, such as nonlinear co-variate effects, overcoming the assumption of linearity in the data.

\begin{equation} \label{eq:GAM}
    E(Y|X_1, X_2,..., X_p) \gets s_0 + \sum_{j=1}^{p} s_j(X_j)
\end{equation}

$Y$ is the dependent variable (the prediction goal), and $E(Y|X_1, X_2,..., X_p)$ represents the function that connects the expected value to the predictor variables $X_1,..., X_p$. $s_0$ is a constant term, and the Backfitting algorithm estimates it.\footnote{The Backfiting algorithm is an iterative procedure to estimate the nonlinear effects of a GAM, enabling the model to handle the nonlinear relationship between the predictor variables and the dependent variable\cite{david2003backfitting,hastie1987generalized}.} The terms $s_1(X_1),..., s_1(X_p)$ denote smooth, nonparametric functions~\cite{larsen2015gam}.\footnote{It is worth mentioning that nonparametric functions, as opposed to their parametric counterparts, derive their shape exclusively from empirical data without reliance on a limited set of predefined parameters~\cite{larsen2015gam}.} Moreover, several smoothers exist in the literature. In this work, GAMs use the smoothing splines~\cite{baayen2020introduction} following the implementation in the R programming language~\cite{r2023programming}.\footnote{R is a programming language designed for statistical computing. It is important in our context because it provides extensive data analysis and presentation tools~\cite{r2023programming}.} The individual variables are separated by the ``+'' operator, indicating the additive characteristic of the model. 

Lastly, to mitigate the potential influence of outlier participants with considerably divergent views on hate speech, we rely on GAMs' ability to manage response data rated on the Likert scale while maintaining robustness against the perturbing impact of outliers in deriving conclusions. GAMs accomplish this by assuming that outliers exhibit distinct behavior, connecting the associated uncertainty with wide confidence intervals~\cite{baayen2020introduction}. 

\section{Study Design}
\label{sec:Design}
Our user study is designed to assess three interpretability layouts based on participants' responses. Thus, we directly ask participants how their views change when presented with interpretability results from each of the three layouts. To achieve this, we create a questionnaire for each layout, consisting of 20 tuples, each comprising two questions. The questionnaire begins by just presenting a sentence (without interpretability data). After that, participants reevaluate the same sentence. However, this time with information from the interpretability tool, as those in Figure~\ref{fig:UserStudyAllSettings}. Concretely, the participants assess each sentence twice, one with no interpretability data and the other with interpretability data.\footnote{Response data and code at:~\url{https://bitbucket.org/thiago-phd/user_study}.} 

The study design comprises both within-subject and between-subject settings. Within-subject refers to assessing whether the information of a given interpretability layout influences the participant's classification of a sentence as hate speech or not (compared to the baseline with no interpretability). This experiment setting considers the same group of participants. In contrast, the between-subject component of the study assesses whether one of the interpretability layouts has a more significant impact on participants' views than another, which involves using different groups of participants. We randomly assign participants to one of three groups. Each group evaluates a different interpretability layout\footnote{We do this to ensure that participants do not use information from one interpretability layout when answering the questionnaire related to another layout.} and consists of 38 participants, totaling 114, and all groups share the same demographic distribution (gender and ethnicity). In adherence to ethical considerations associated with human studies, the experiment protocol was subject to review by the Universitat Autònoma de Barcelona ethics committee.

\paragraph{Power Analysis.}We perform this calculation to determine the appropriate group size, which requires defining the following factors: 1) statistical power, the probability that the statistical test will reject the null hypothesis and obtain a statistically significant result~\cite{cohen1992statistical}; 2) effect size, the quantitative indicator that measures the magnitude of the difference between distributions (i.e., the impact of a layout) when conducting a statistical analysis~\cite{brysbaert2018power}; 3) significance level ($\alpha$), the threshold to define the presence of statistically significant difference~\cite{cohen1992statistical}; and 4) Cohen's $d$ value, the measure of effect size~\cite{brysbaert2018power}.

This power analysis considers the effect size measured with Cohen's $d$ value~\cite{cumming2014new} obtained from the investigation by~\cite{yang2020visual}. This work is particularly relevant to our user study because it addresses a similar experiment context, i.e., the investigation of interpretability in ML models. We aim to achieve a statistical power of $0.95$ while maintaining a significant level of $\alpha=0.05$, for which we set Cohen's $d$ value to $0.84$~\cite{yang2020visual}. Moreover, we use the two-tailed T-test method. This statistical hypothesis test examines the existence of a significant difference between two distributions in both directions, whether the difference indicates an increase or decrease in the influence of an interpretability layout~\cite{park2015hypothesis}. Consequently, following this calculation, we get to 38 participants per group. Lastly, it is worth noting that our required number of participants is consistent with the sizes employed in other studies in the literature~\cite{arora2022explain,chu2020visual,schuff2022human,yang2020visual}, aligning with established protocols in the field and further supporting our decision. 

\paragraph{Questionnaire.}Each participant must evaluate 20 sentences, ten for each of the two hate speech classes in this study, ``Misogyny'' and ``Racism.'' It is worth noting that the participants across the three distinct groups see the same 20 sentences. The difference between the groups lies in their usage of different interpretability layouts. To determine the number of sentences in the study, we consider three main points: 1) ensure that the task completion time is not excessively long to avoid potential fatigue effects~\cite{zhang2018understanding}; 2) limit the number of hate-speech sentences evaluated by the participants to avoid any negative impact (e.g., anxiety, stress, and sadness) on their mental health~\cite{karunakaran2019testing,steiger2021psychological}; and 3) align the number of sentences with previous studies that achieved significant results~\cite{chu2020visual,de2003effects,yang2020visual}. 

The selected sentences are balanced to ensure that both classes are represented equally, choosing ten distinct sentences for each of the two respective classes. To avoid any effect grammatical mistakes might have on participants' evaluations, we apply grammatical corrections when needed,\footnote{For instance, ``fuck y all N-Word'' was corrected to ``Fuck you all N-Word.''} without changing a sentence's meaning. We obtained 40 answers from each participant during the user study. This includes two answers for each sentence (one with no interpretability, the other with interpretability information). In both cases, participants rate the classification of each sentence on a 7-point Likert scale. Across all three layouts, we collected 1,520 answers per layout, with 38 participants providing 40 answers each (4,560 answers for the entire study). The questionnaire presents the text sentences randomly to address potential carry-over effects.

\paragraph{Statistical Analysis.}Our user study evaluates two hate speech classes, ``Misogyny'' and ``Racism.'' The reason behind this choice is to maintain the study concise and allow us to control the variables in the experiment, which we accomplish by accounting for demographic factors (gender and ethnicity) in our statistical analysis. Specifically, this involves limiting gender identification to \emph{male} and \emph{female}, and ethnicity identification to \emph{black} and \emph{white}.\footnote{We note the existence of other gender and ethnicities identities~\cite{pewresearch2020,pega2015case}. However, aiming to keep the study concise and statistically sound, we limit the options to only two for each class.} This approach takes into consideration that gender may influence perceptions of misogynistic behavior, with women reportedly identifying instances of this behavior more frequently than men~\cite{kirkman2020just}. Furthermore, black and white individuals perceive racism differently, with distinct perceptions about attention to racial issues, inequality, and violence targeting black people~\cite{pewresearch2020}. Including these demographic variables in our statistical analysis enables an evaluation of whether participant characteristics influence hate speech classification. In other words, our statistical analysis aims to isolate the variable of interest (interpretability layout) while simultaneously accounting for potential confounding variables (gender and ethnicity). By adopting this approach, we preserve the conclusions of our statistical analysis, ensuring that confounding variables do not incorrectly impact our results about the true relationship between the variables of interest.

Our GAM (Section~\ref{subSec:GAM}), employed to estimate the participants' classification ratings, comprises fixed effects, including interpretability layouts (also referred to as treatment) and demographic aspects (gender and ethnicity), and random effects, such as the participants and text sentences. Equation~\ref{eq:UserStudySpecificGAM} describes the specific model for this study.

\begin{equation} \label{eq:UserStudySpecificGAM}
    y_{target} \gets \beta_0 + \beta_{treat} \times x_{treat} + \beta_{dem} \times x_{dem} + \alpha_{user} \times x_{user} + \alpha_{sent} \times x_{sent}
\end{equation}

$y_{target}$ refers to participants' answers on the Likert scale, while $x_{user}$ and $x_{sent}$ refer to random effects corresponding to individual participants and text sentences, respectively. Regarding fixed effects, $x_{dem}$ represents demographic aspects, and $x_{int}$ represents the different interpretability layouts. Additionally, $\beta_0$ represents the intercept,\footnote{The intercept is a term in the model to represent the estimated value of the dependent variable when the values of the independent variables are 0.} while $\beta_{treat}$ and $\beta_{dem}$ refer to the weight for the interpretability layouts and demographic factors, respectively. $\alpha_{user}$ and $\alpha_{sent}$ address the weights for the random effects, considering participant-specific and sentence-specific effects, respectively.   

\paragraph{Participant Recruitment.}We recruit participants from the Prolific crowdsourcing platform~\cite{prolific2023platform}, which complies with the European GDPR data protection and treatment framework~\cite{voigt2017eu}. To complete the study, participants must read the research goal, give their consent, and answer the questionnaire. We select this platform because it covers the following requirements: 1) easy integration with external forms (e.g., Alchemer) and easy management of the study lifecycle (i.e., the response collection stages); 2) our previous experience with the platform; and 3) as presented in the Fairwork Cloudwork report~\cite{fairwork2022}, it implements policies aiming to improve work conditions by mitigating precarity and overwork while allowing researchers to directly set the amount paid to each worker.

We set the payment to £13,00 an hour.\footnote{£ refers to Pound Sterling.} Since the maximum time of the experiment is 30 minutes, each participant is paid £6,50, regardless of whether they use all this time or not (within this 30-minute window). Additionally, our consent form informs participants that if they decide to leave the study without completing it, they will receive compensation according to the time spent on the questionnaire (based on the 
£13,00 per hour rate). Following ethical guidelines defined in the platform~\cite{prolific2023payment}, we set payment to all participants equally, avoiding negative psychological effects on crowdsourcing workers due to payment based on performance conditions (e.g., getting the ``right answers'')~\cite{sayre2023costs}. For questionnaires, the platform recommends payments of £9,00 per hour. However, due to the nature of our sentences (hate speech), we add £4 per hour (44\% increase).

To ensure the quality of answers, we use attention checks to measure the participant's attention while answering our questionnaire. Six of the 114 initially recruited participants failed these attention checks. Thus, we recruited six additional people to achieve our desired number of participants. 

\section{Results}
\label{sec:Results}

\subsection{Within-Subject}
\label{subSec:Within}
The within-subject part of our user experiments presents three sets of results, one for each interpretability layout. Our analysis compares the interpretability layouts to the baseline to obtain these results. This comparison allows us to understand whether interpretability layouts influence participants' classification of a sentence as a particular class of hate speech, ``Misogyny'' and ``Racism.''


The possible values in our GAM implementation described in Equation~\ref{eq:UserStudySpecificGAM} are: 
\begin{itemize}
    \item Treatment - a) local interpretability; b) sum of relevance scores (``list''); and c) combined approach.
    \item Gender - a) male; and b) female.
    \item Ethnicity - a) black; and b) white.
    \item Random Factors (users and sentences) - the possible values in these variables are the individual participants of the user study (38 for each interpretability layout) and the 20 sentences we present to these participants, respectively.
\end{itemize}

Table~\ref{tab:SummarizedParametricCoefficients} provides the results for the local interpretability layout compared to the baseline. Results indicate that none of the factors significantly influence participants' classifications, as corroborated by the Pr($>|z|$) values. In other words, the participants' views about hate speech are not affected by the interpretability layout, nor their gender or ethnicity. Our findings indicate that the individual sentences and users influence the classification ratings, as demonstrated in Table~\ref{tab:SummarizedSignificanceSmooth}. The reasons for this conclusion are twofold: 1) some sentences might exhibit more explicit hate speech than others. Thus, a sentence containing explicit racist terms may prompt participants to assign higher ratings on the Likert scale than sentences containing subtle expressions of racism; and 2) some participants may have divergent perspectives regarding the severity of hate speech. Thus, these participants may assign higher ratings on the Likert scale than others, reflecting a greater inclination towards classifying hate speech with severity. 

\begin{table}[t!]
\captionsetup{font=small}
	\centering
    \begin{tabular}{l|cccc}
        Factors & Estimate & Std. Error & $z$-value & Pr($>|z|$)\\
		\hline
		\textbf{Intercept} & 4.0429 & 0.6049 & 6.683 & $2.33e^{-11}$ \\
		\hline
		\textbf{Treatment - Local} & -0.2243 & 0.6324 & -0.355 & 0.723 \\
		\hline
		\textbf{Gender - Male} & -0.5905 & 0.4803 & -1.230 & 0.219 \\
            \hline
            \textbf{Ethnicity - White} & -0.3223 & 0.4802 & -0.671 & 0.502 \\
    \end{tabular}
\caption[The fixed terms for the local interpretability layout experiment]{The fixed terms for the experiment considering the local interpretability layout. The parametric coefficients have values for Estimate, Standard Error, z-value, and Pr($>|z|$). The $z$-value relates to the estimation's mean, representing the number of standard deviations from this mean. Lastly, the Pr($>|z|$) column depicts the p-value for the coefficients. Specifically, it indicates the probability of obtaining a value of $z$ bigger than our calculated absolute $z$-value.}
\label{tab:SummarizedParametricCoefficients}
\end{table}

\begin{table}[t!]
\captionsetup{font=small}
	\centering
    \begin{tabular}{l|cccc}
        Smooth Terms & EDF & Ref.df & Chi.sq & P-value\\
		\hline
		\textbf{User Id} & 33.19 & 35 & 1374 & $< 2 e^{-16}$ \\
		\hline
		\textbf{Sentence Id} & 36.83 & 38 & 2367 & $< 2 e^{-16}$ \\
    \end{tabular}
\caption[The smooth terms for the local interpretability layout experiment]{Approximate significance of smooth terms for the experiment considering the local interpretability layout. With Effective Degrees of Freedom (EDF), Reference Degrees of Freedom (Ref.df), Chi.sq, and p-value. EDF represents the complexity of the smooth. For instance, an EDF of 1 indicates a straight line. Higher EDFs represent more wiggly curves. The Ref.df column contains the maximum degrees of freedom for each term used in calculating the p-value. The Chi.sq represents the test statistic to determine the significance of the smooth. Lastly, the p-value is the result of the test.}\label{tab:SummarizedSignificanceSmooth}
\end{table}

Regarding the interpretability layout that uses the sum of relevance scores, results are detailed in Tables~\ref{tab:ListSummarizedParametricCoefficients} and~\ref{tab:ListSummarizedSignificanceSmooth}. Like the local interpretability case, the results indicate that no fixed terms (treatment, gender, and ethnicity) influence participants' classification of hate speech sentences. The individual users and sentences are the only effects influencing the classification ratings.  

\begin{table}[t!]
\captionsetup{font=small}
	\centering
    \begin{tabular}{l|cccc}
        Factors & Estimate & Std. Error & $z$ Value & Pr($>|z|$)\\
		\hline
		\textbf{Intercept} & 4.8605 & 0.6170 & 7.88 & $3.3e^{-15}$ \\
		\hline
		\textbf{Treatment - List} & -0.0174 & 0.5984 & -0.03 & 0.98 \\
		\hline
		\textbf{Gender - Male} & -0.6228 & 0.5121 & -1.22 & 0.22 \\
            \hline
            \textbf{Ethnicity - White} & -0.1098 & 0.5121 & -0.21 & 0.83 \\
    \end{tabular}
\caption[The fixed terms for the sum of score interpretability layout experiment]{The fixed terms for the experiment considering the sum of relevance scores interpretability layout (``list'').}
\label{tab:ListSummarizedParametricCoefficients}
\end{table}

\begin{table}[t!]
\captionsetup{font=small}
	\centering
    \begin{tabular}{l|cccc}
        Smooth Terms & EDF & Ref.df & Chi.sq & P-value\\
		\hline
		\textbf{User Id} & 33.4 & 35 & 1623 & $< 2 e^{-16}$ \\
		\hline
		\textbf{Sentence Id} & 36.7 & 38 & 2426 & $< 2 e^{-16}$ \\
    \end{tabular}
\caption[The smooth terms for the sum of relevance scores interpretability layout experiment]{Approximate significance of smooth terms for the experiment considering the sum of relevance scores interpretability layout (``list'').}\label{tab:ListSummarizedSignificanceSmooth}
\end{table}

The set of results for the combined interpretability layout is the last part of the within-subject experiments. Tables~\ref{tab:CombinedSummarizedParametricCoefficients} and~\ref{tab:CombinedSummarizedSignificanceSmooth} detail the corresponding statistical values. As in the previous interpretability layout cases, no fixed term influences participants' classification of violating behavior, with only the specific user and sentence being relevant.

\begin{table}[t!]
\captionsetup{font=small}
	\centering
    \begin{tabular}{l|cccc}
        Factors & Estimate & Std. Error & $z$ Value & Pr($>|z|$)\\
		\hline
		\textbf{Intercept} & 4.1039 & 0.6110 & 6.72 & $1.9e^{-11}$ \\
		\hline
		\textbf{Treatment - Combined} & -0.0202 & 0.6620 & -0.03 & 0.98 \\
		\hline
		\textbf{Gender - Male} & -0.4544 & 0.4490 & -1.01 & 0.31 \\
            \hline
            \textbf{Ethnicity - White} & -0.5741 & 0.4490 & -1.28 & 0.20 \\
    \end{tabular}
\caption[The fixed terms for the combined interpretability layout experiment]{Fixed terms for the experiment considering the combined interpretability layout.}\label{tab:CombinedSummarizedParametricCoefficients}
\end{table}

\begin{table}[t!]
\captionsetup{font=small}
	\centering
    \begin{tabular}{l|cccc}
        Smooth Terms & EDF & Ref.df & Chi.sq & P-value\\
		\hline
		\textbf{User Id} & 32.9 & 35 & 983 & $< 2 e^{-16}$ \\
		\hline
		\textbf{Sentence Id} & 36.7 & 38 & 1718 & $< 2 e^{-16}$ \\
    \end{tabular}
\caption[The smooth terms for the combined interpretability layout]{Approximate significance of smooth terms for the experiment considering the combined interpretability layout.}\label{tab:CombinedSummarizedSignificanceSmooth}
\end{table}

In summary, our findings demonstrate that no interpretability layout influences participants' views on misogyny and racism. Furthermore, the demographic factors under investigation, specifically gender and ethnicity, do not yield any significant influence on participants' assessments. We argue that participants start with a given view of what constitutes hate speech when they answer the questionnaire, or our model's output already aligns with their classification of hate speech. Thus, any information provided by the interpretability layouts fails to change the participant's rating. This conclusion is supported by our qualitative analysis (Section~\ref{subSec:Quali}), with comments highlighting participants' reliance on their own view of hate speech when assessing the sentences.

\subsection{Between-Subject}
\label{subSec:Between}
In addition to executing a within-subject analysis, we aim to investigate direct differences between interpretability layouts, which enables us to understand if an interpretability layout asserts more influence on participants than others. 

The results in Table~\ref{tab:BetweenSummarizedParametricCoefficients} indicate that no particular interpretability layout influences participants' responses more than the others. Furthermore, the other fixed terms, gender and ethnicity, similarly do not affect the ratings across the different layouts. In the context of the random effects, Table~\ref{tab:BetweenSummarizedSignificanceSmooth} demonstrates that the differences in classification ratings can be attributed to the specific user and sentence, aligning with the findings in our within-subject analysis.

\begin{table}[t!]
\captionsetup{font=small}
	\centering
    \begin{tabular}{l|cccc}
        Factors & Estimate & Std. Error & $z$ Value & Pr($>|z|$)\\
		\hline
		\textbf{Intercept} & 4.1595 & 0.5041  & 8.25 & $2e^{-16}$ \\
		\hline
		\textbf{Treatment - List} & -0.0202  & 0.3267 & -0.06 & 0.95 \\
		\hline
		\textbf{Treatment - Local} & -0.2558 & 0.3265 & -0.78 & 0.43 \\
            \hline
            \textbf{Gender - Male} & -0.4674 & 0.2666 & -1.75 & 0.08 \\
            \hline
            \textbf{Ethnicity - White} & -0.3523 & 0.2666 & -1.32 & 0.19 \\
    \end{tabular}
\caption[The fixed terms for the between-subject experiment]{The fixed terms for the between-subject experiment, comparison of three interpretability layouts.}\label{tab:BetweenSummarizedParametricCoefficients}
\end{table}

\begin{table}[t!]
\captionsetup{font=small}
	\centering
    \begin{tabular}{l|cccc}
        Smooth Terms & EDF & Ref.df & Chi.sq & P-value\\
		\hline
		\textbf{User Id} & 97.4 & 109 & 2918 & $< 2 e^{-16}$ \\
		\hline
		\textbf{Sentence Id} & 18.8 & 19 & 2071 & $< 2 e^{-16}$ \\
    \end{tabular}
\caption[The smooth terms for the between-subject experiment]{Approximate significance of smooth terms considering the between-subject experiment.}\label{tab:BetweenSummarizedSignificanceSmooth}
\end{table}

\subsection{Qualitative Evaluation}
\label{subSec:Quali}

The previous two sections address the statistical analysis of our user study. Here, we investigate the qualitative aspect of this experiment, aiming to gain insights from participants' comments regarding what they consider relevant while evaluating the sentences. Although most of the 114 comments do not convey any relevant message to our investigation, we shall discuss the comments that complement our statistical analysis in the rest of this section. Specifically, we analyze comments that fit into three categories: 1) familiarity with the task, comprising four comments; 2) questionnaire design, comprising five comments; and 3) model accuracy, comprising four comments. Next, we discuss the most representative instances within each category.

\paragraph{Familiarity with the Task.}An interesting insight regarding participants' perception of how they used information from the interpretability layouts and their familiarity with the hate speech classification task is present in the following comment: ``\emph{I did not find that the AI-generated significance values affected my perceptions of whether statements were misogynistic or racist, but rather confirmed my thoughts, with a couple of exceptions that read more benign (something you'd read in a history book, e.g., the `royal strippers' example) that the AI represented were more than I believed.}'' This observation supports our reasoning that, in the context of hate speech - a common type of violation that people are exposed to in online interactions, participants tend to have a clear view of what constitutes hate speech before they are presented with interpretability information. Concretely, this comment highlights two scenarios: 1) our interpretability information already aligns with what participants expect of hate speech, i.e., the ML model correctly learns these kinds of violations; 2) if the interpretability information does not align with participants' definition of hate speech, then it is difficult to change people's views on what they consider hate speech when they evaluate the sentences present in our study (such as the presence of sentences with ``royal strippers''). The following comment provides further insight into participants' familiarity with the hate speech task: ``\emph{...Maybe it went so quick [the time needed to complete the questionnaire] because the language used was very obvious for me in choosing an answer.}'' This observation highlights that the participant has a clear view of what constitutes hate speech. Moreover, the sentences evaluated in our study clearly contain terms related to hate speech that people are familiar with.

\paragraph{Questionnaire Design.}Another relevant observation relates to the questionnaire design, as highlighted in the following comment: ``\emph{This survey was a little unpleasant, as stated in the consent form. Thank you for not making this survey any longer than it is now.}'' Our initial goal was to maintain survey conciseness, strictly including only the required number of sentences to achieve our predetermined statistical power. The comment's observation supports our initial decision, which is especially relevant since we are handling offensive language that could be hurtful for the readers, especially when they are numerous. As such, we strongly recommend that future research tackling sensitive topics like hate speech follow in our footsteps and carefully consider the number of violations presented in questionnaires, conducting statistical tests and comparing with existing literature to avoid exposing participants to sentences that may adversely affect their mental health. In this context, our work provides further metrics based on statistical analysis to assist in future questionnaire design processes.

\paragraph{Model Accuracy.}Participants expressed concerns about the relevant terms identified by our interpretability approach. The following comment illustrates this concern: ``\emph{The highlighting seemed strangely unfocused in some instances...}'' This observation highlights an important aspect in the ML field: the discrepancy between the ML model and human perception. Thus, we argue that if interpretability is not used to influence or correct a user's behavior, such layouts can be useful to identify these discrepancies and trigger user feedback that would help correct the outputs of the ML model. Concretely, the existence of a contrast between participants' views and the model's output supports incorporating interpretability layouts to assist the evaluation of the model's behavior beyond conventional performance metrics (e.g., recall, F1-score, and precision). Moreover, given that our previous works~\cite{freitas2023cross} focused on domains characterized by small labeled datasets, there might be instances where the model fails to capture terms commonly associated with offensive content. Consequently, this may lead to the model's inability to identify specific sentences as hate speech, impacting its accuracy. For instance, the following comment is a representative case: ``\emph{The AI is unaware of racial nuances behind certain phrases such as `big lipped' which was not even highlighted.}'' This observation illustrates potential limitations in the detection capabilities of our model, especially when the model must handle terms that have limited representation in our training dataset (for instance, one occurrence). Integrating interpretability enhances the transparency of ML models by enabling people to understand the relevant terms that these models consider. This is particularly significant considering the dynamic nature of online interactions, where new sentences expressing hate speech emerge over time within a single community or as we incorporate data from diverse communities.

\section{Literature Review}
\label{sec:Literature}
The survey in~\cite{rong2022towards} offers important guidelines for user studies in Explainable Artificial Intelligence (XAI), exploring interesting works in this domain. Their discussion regarding the effectiveness of explanations in increasing participant trust and usability of ML models is particularly relevant to our research. However, it is important to note that existing literature presents studies with different results. Consequently, we emphasize the importance of our work in providing additional empirical evidence in the field of assessing the impact of interpretability. 

Schuff et al.~\cite{schuff2022human} investigate how people understand and interpret explanations provided by the word salience in text sentences. Their methodology shares similarities with the information in Figure~\ref{fig:UserStudyLocalInterpretability}, where salience indicates the terms that most influence the model's decision-making process. They execute a study to evaluate the impact of different factors (e.g., word frequency, word length, display index) on a participant's interpretation of the explanation. Like our study, they use crowdsourcing workers and employ a GAM-based approach to analyze the data. However, certain differences are important to note. First, they compare bar charts to heatmap-based salience visualizations, while we compare salience (local interpretability), list, and a combination of both. Second, their study focuses on textual data elements as factors in the model, while we incorporate demographic variables as influencing factors. Lastly, instead of asking about the participants' views regarding the model's classification, they inquire how relevant a specific word is to the model's output (also on a Likert scale).

Arora et al.~\cite{arora2022explain} explore a model that must differentiate between authentic and fake hotel reviews. Their study asks participants to simulate the model's behavior on new reviews after having access to explanations. The goal is to assess whether interpretability helps humans predict the model output on new instances. Additionally, participants are asked to engage in a manipulation task by editing the review to change the classification output. This step evaluates if participants can lower the model confidence towards the original predicted class (i.e., lower the classification probability). Like our findings demonstrating that interpretability does not influence participants' views, their results indicate that local interpretability fails to improve participants' capacity to replicate the model's decision-making process. However, the authors conclude that participants can affect the model's confidence by leveraging information from the global attributions tool, built using a linear model that emulates a Pre-Trained Language Model. These general explanations provide insights into common input-output associations that the models exploit. Lastly, they employ IG as the interpretability algorithm and use mixed effects models to analyze participants' interactions.

Unlike our work that evaluates interpretability in the scope of textual data, Chu et al.~\cite{chu2020visual} address saliency-based interpretability techniques applied to image data, specifically in the context of age prediction tasks. Their study measures whether effective model explanations enhance human accuracy (i.e., predict age better) while quantifying whether flawed explanations decrease human trust (i.e., quantify the difference between people's answers and the model's output). The findings indicate that different types of explanations do not significantly influence human accuracy or trust, which is similar to our work's conclusion, where different interpretability tools fail to impact participants' views. They use a mixed-effects model to estimate error, measuring the difference between participants' guesses about a person's age and the true label. Also working with image, Alqaraawi et al.~\cite{alqaraawi2020evaluating} find that the saliency-based approach helps participants comprehend specific image features (i.e., the relevant pixels) that contribute to the output of an ML model solving an image classification tasks, such as identifying objects. It is worth noting that their results indicate that this approach is limited in assisting participants in anticipating the model's predictions for new images.

Similar to our combined layout, Radensky et al.~\cite{radensky2022exploring} conduct a user study to assess the effectiveness of combining local and global information compared to relying solely on a single interpretability layout. Their findings indicate that the combination is better at assisting participants to comprehend how a recommender system can improve. It is worth noting that this result differs from our findings, as in our domain, the combined layout does not present a higher impact compared to the others, with no significant influence on participants' views.

Wang et al.~\cite{wang2021explanations} conduct an investigation that compares different XAI methods in AI-assisted decision-making. Their study describes three properties that should be present in AI explanations to make them helpful from a human perspective: model understanding, uncertainty recognition, and trust calibration. Their findings indicate that the nature of the decision-making task significantly impacts the efficacy of the employed XAI techniques. However, the results also show that when participants lack domain knowledge, the XAI methods do not satisfy the investigated properties. Thus, they recommend building XAI methods specific to these cases, incorporating alternative techniques to provide interpretability information. We note that this recommendation differs from our study regarding our evaluation of two classes of hate speech, a domain where participants generally have some prior knowledge.

Lastly, regarding user studies using crowdsourcing platforms, Shank~\cite{shank2016using} presents concepts, patterns, and suggestions for exploring these platforms in online research. Their review showcases several papers where researchers successfully acquire reliable and quality data in diverse contexts such as psychology and economics. Additionally, one essential feature of these platforms relevant to our study is their ability to easily target specific demographic characteristics within a particular population. In our use case, we aim to target participants based on ethnicity and gender.

\section{Conclusion}
\label{sec:Conclusion}
This paper presented a user study to assess the potential influence of three interpretability layouts when participants evaluate hate speech in text, focusing on the ``Misogyny'' and ``Racism'' classes. Concretely, this work answered the following question: can interpretability layouts influence human perception of offensive sentences? As such, with statistical and qualitative analysis of questionnaire responses, we provided empirical evidence in the context of ML interpretability within online communities.

The three interpretability layouts considered in this work comprised local, the sum of relevance scores (``list''), and a combined approach, as illustrated in Figure~\ref{fig:UserStudyAllSettings}. The Integrated Gradients (IG, Section~\ref{subSec:IG}) algorithm provided interpretability data for the study. We used an ML model~\cite{freitas2023cross} that combines adapters with Pre-Trained Language Models to obtain the interpretability data. 

Moreover, we ensured that different factors (gender and ethnicity) did not incorrectly impact our results by considering this information in our statistical analysis. The literature~\cite{pewresearch2020,kirkman2020just} indicates that gender and ethnicity might influence perceptions of misogynistic and racist behavior, respectively. Thus, including these demographic characteristics was motivated by our focus on the two hate speech classes: ``Misogyny'' and ``Racism.'' 

This user study comprised both within-subject and between-subject analyses. Additionally, it covered a qualitative evaluation using participants' comments. Participants were randomly allocated into three groups, corresponding to the three distinct interpretability layouts. A power analysis determined a sample size of 38 participants per group. Each participant evaluated 20 text sentences, equally divided between the two violation classes, performing two evaluations per sentence: one without interpretability data and one with data from one of the three layouts. This study design yielded 40 responses per participant, collected on a 7-point Likert scale. We employed the Generalized Additive Model (GAM) (Section~\ref{subSec:GAM}) to estimate participants' ratings and analyze the influence of interpretability layouts, gender, ethnicity, individual sentences, and participants on the responses (Section~\ref{sec:Results}). 

Results of our statistical analysis indicated that none of the interpretability layouts influenced participants' views on classifying hate speech in the "Misogyny" and "Racism" classes. Furthermore, no single layout had a more significant impact than the others. Qualitative analysis of participants' comments leveraged our comprehension of these results. Two key points explain why interpretability layouts did not significantly influence participants' views: familiarity with hate speech, the presence of explicit terms, and the alignment between the ML model's output and participants' views, where the interpretability layout highlighted relevant terms similar to those identified by participants. Comments on the model's accuracy highlighted the importance of ML interpretability and feedback mechanisms. Specifically, interpretability addresses transparency by allowing participants to understand the relevant terms the model considers, while feedback enables correction of misalignment between participants' views and the model's output. This is especially important in contexts where new instances of hate speech may emerge over time. Moreover, interpretability layouts can provide insights into evaluating the ML model's behavior beyond traditional performance metrics (e.g., recall, F1-score, and precision), which is especially relevant in scenarios where understanding the rationale behind an ML model's decision is essential for optimal implementation. 

Future work shall investigate other norm violations unrelated to hate speech. This aims to address the limitation of our user study in evaluating a task (classification of a text sentence as racist or misogynistic) already familiar to participants. Our goal is to investigate tasks in different domains, such as the interaction of people during online meetings, where norms might regulate the meeting's duration, the volume of messages exchanged, and when participants should interact. In this context, interpretability information describing expected behavior within a particular online community may assist novice members in understanding and adhering to established norms. For instance, if the volume of message exchange exceeds established norms, interpretability can provide information regarding the expected volume in this community. Again, this information can contribute to a future feedback elicitation process, where new definitions of the elements (i.e., the new expected volume of message exchange) that comprise violating behavior in this domain may emerge. A new user study can help assess whether community members use interpretability information to adapt their behavior (without human interventions) in online meetings. We believe the investigation in this use case has the potential to yield different results compared to the hate speech use case, as participants may not have established opinions about what constitutes a violation in this context. Additionally, when they have such notions, expected behavior may vary greatly depending on the individual views. As such, our assumption is that participants will demonstrate a higher tendency to adapt to the definitions provided by the interpretability tool.

\newpage
%
%
%
\bibliographystyle{splncs04}
\bibliography{references}

\end{document}